\documentclass[letterpaper,twocolumn,english,aps,prx,floatfix,superscriptaddress,nofootinbib]{revtex4-1}
\usepackage{lmodern}

\usepackage{amsmath}
\usepackage{amssymb}
\usepackage{graphicx}
\usepackage{esint}
\usepackage{setspace}

\usepackage{lipsum}
\makeatletter

\pdfpageheight\paperheight
\pdfpagewidth\paperwidth

\@ifundefined{textcolor}{}
{%
 \definecolor{BLACK}{gray}{0}
 \definecolor{WHITE}{gray}{1}
 \definecolor{RED}{rgb}{1,0,0}
 \definecolor{GREEN}{rgb}{0,1,0}
 \definecolor{BLUE}{rgb}{0,0,1}
 \definecolor{CYAN}{cmyk}{1,0,0,0}
 \definecolor{MAGENTA}{cmyk}{0,1,0,0}
 \definecolor{YELLOW}{cmyk}{0,0,1,0}
}

\renewcommand{\[}{\begin{equation}}
\renewcommand{\]}{\end{equation}}

\usepackage{babel}

\makeatother

\usepackage[titles]{tocloft}

\usepackage{enumitem}

\begin{document}
\global\long\def\avg#1{\langle#1\rangle}

\global\long\def\pr{\prime}

\global\long\def\dg{\dagger}

\global\long\def\ket#1{|#1\rangle}

\global\long\def\bra#1{\langle#1|}

\global\long\def\proj#1#2{|#1\rangle\langle#2|}

\global\long\def\inner#1#2{\langle#1|#2\rangle}

\global\long\def\tr{\mathrm{tr}}

\global\long\def\pd#1#2{\frac{\partial#1}{\partial#2}}

\global\long\def\spd#1#2{\frac{\partial^{2}#1}{\partial#2^{2}}}

\global\long\def\der#1#2{\frac{d#1}{d#2}}

\global\long\def\im{\imath}

\global\long\def\onlinecite#1{\cite{#1}}

\global\long\def\c{\mathcal{C}}

\global\long\def\r{\mathcal{R}}

\global\long\def\e{\mathcal{E}}

\global\long\def\t{\mathcal{T}}

\global\long\def\s{\mathcal{S}}

\global\long\def\p{\mathcal{P}}

\title{Metastable morphological states of catalytic nanoparticles}

\author{Pin Ann Lin}

\affiliation{Center for Nanoscale Science and Technology, National Institute of
Standards and Technology, Gaithersburg, Maryland 20899, USA}

\affiliation{Maryland NanoCenter, University of Maryland, College Park, Maryland,
USA. }

\author{Bharath Natarajan}

\affiliation{Materials Measurement Laboratory, National Institute of Standards
and Technology Gaithersburg, MD 20899 USA}

\author{Michael Zwolak}
\email{michael.zwolak@nist.gov}

\selectlanguage{english}%

\affiliation{Center for Nanoscale Science and Technology, National Institute of
Standards and Technology, Gaithersburg, Maryland 20899, USA}

\author{Renu Sharma}
\email{renu.sharma@nist.gov}

\selectlanguage{english}%

\affiliation{Center for Nanoscale Science and Technology, National Institute of
Standards and Technology, Gaithersburg, Maryland 20899, USA}
\begin{abstract}
During the catalytic synthesis of graphene, nanotubes, fibers, and
other nanostructures, many intriguing phenomena occur, such as phase
separation, precipitation, and analogs of capillary action. We demonstrate
that catalytic nanoparticles display metastable states that influence
growth, reminiscent of some protein ensembles \emph{in vivo}. As a
carbon nanostructure grows, the nanoparticle elongates due to an energetically
favorable metal-carbon interaction that overrides the surface energy
increase of the metal. The formation of subsequent nested tubes, however,
drives up the particle\textquoteright s free energy, but the particle
remains trapped until an accessible free energy surface allows it
to exit the tube. During this time, the nanoparticle continues to
catalyze tube growth internally within the nested structure. This
nonequilibrium thermodynamic cycle of elongation and retraction is
heavily influenced by tapering of the structure, which, ultimately,
determines the final product and catalyst lifetime. Our results provide
a unifying framework to interpret similar phenomena for other catalytic
reactions, such as during CO oxidation, and suggest routes to the
practical optimization of such processes.
\end{abstract}

\maketitle

Advances in catalytic growth \textendash{} especially chemical vapor
deposition {[}1,2{]} \textendash{} of carbon nanostructures are paving
the way for their ubiquitous application in technologies. Their exceptional
thermal, electronic, and mechanical properties {[}3\textendash 9{]}
are primarily determined by their structure, which is in turn determined
by their growth conditions (composition, size, and shape of the catalyst,
support, temperature, etc.). However, the product is invariably a
mixture of all possible carbon nanostructures instead of a single
phase with the desired properties for a specific application. Understanding
the chemical and morphological evolution of catalyst particles during
nanostructure nucleation and growth will help determine conditions
needed for controlled synthesis. 

Recently, environmental transmission electron microscope (ETEM) has
been successfully employed to follow the nucleation and growth of
various carbon nanostructures on catalytic particles {[}10\textendash 12{]}.
One perplexing observation is the appearance of morphological changes
of the metal catalysts {[}10,13,14{]}, first reported by Helveg et
al.~{[}10{]} using ETEM imaging. In this Letter, we provide a thermodynamic
explanation of the elongation-retraction cycle of catalytic nanoparticles
and compare with high resolution ETEM videos of bamboo-like carbon
nanotube (BCNT) growth on nickel nanoparticles supported on SiO$_{2}$,
see the Supplemental Information (SI) for experimental details. The
elongation-retraction cycle occurs for different metal catalysts and
conditions, as well as for other structures, such as boron nitride
nanotubes. Moreover, our model should be extendable to other catalytic
processes where morphological changes occur, such as surface oscillations
in Pt nanoparticles during catalytic oxidation of CO {[}15{]}.

\begin{figure*}
\begin{centering}
\includegraphics[width=1\textwidth]{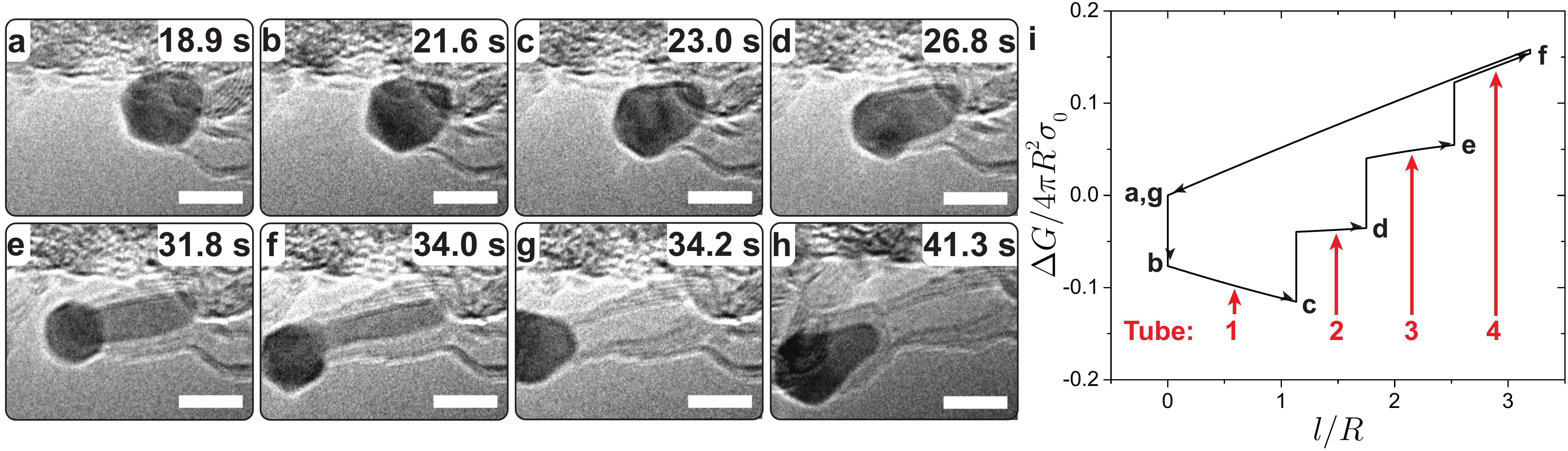}
\par\end{centering}
\caption{\label{fig:RepCycle}Elongation-retraction cycle of a catalytic nanoparticle.
(a-c) An approximately spherical particle first nucleates a nanotube
cap, which subsequently starts to grow into a tube. When the tube
radius is above a threshold value, the particle is favored to elongate
with it (here, into a ``pear shape''). (d-e) The formation of nested
tubes performs work on the particle by decreasing its radius in the
elongated region. (f-g) This eventually results in retraction via
a rapid diffusional process. (h) The process repeats with different
elongation/times/number of tubes for each cycle (see Movie S1 and
Fig. S3). All scale bars are 5 nm. (i) The Gibbs free energy change,
$\Delta G$, of an idealized version of the morphological cycle a$\Rightarrow$g
in units of the surface free energy of the original particle of radius
$R$ and surface energy density $\sigma_{0}$. The morphology shows
many steps of constant tube radius (lateral lines) and constant tube
length (vertical lines). The addition of each nested tube drives the
particle up in energy via work performed on it by the carbon nanotube
growth (the combined nanoparticle, carbon system \textendash{} i.e.,
including the very favorable formation free energy of the nanotube
\textendash{} will have total free energy running downhill). As the
particle\textquoteright s free energy continues to increase, the particle
will eventually find a downhill free energy path to exit the tube.
Here, the particle retracts from the fourth tube (see Fig. S3d,e),
restructuring its tip in the process. An earlier retraction, e.g.,
from the third tube, would require overcoming a significantly larger
free energy barrier.}
\end{figure*}

Figure \ref{fig:RepCycle}a-h shows an elongation-retraction cycle
of nickel growing a BCNT (see also Movie S1). After nucleation, the
metal-carbon interaction makes it energetically favorable for the
catalyst particle to deform with the elongating tube so long as the
radius is above a threshold value. However, as further tubes form
internally to make a multi-walled structure, eventually the elongated
state is unfavorable. The particle, though, remains pinned in the
structure. Only when the innermost tube is both small and long enough
will the nanoparticle be \textquotedblleft released\textquotedblright{}
and the process will repeat. 

This morphological change is energetically costly in terms of the
particle\textquoteright s surface energy (a change of 100\textquoteright s
of eV as the particle deviates from a spherical form), but the metal-carbon
interaction compensates for this: The elongation results in a change
in the metal surface area of $\Delta A$, with an energy penalty $\Delta A\cdot\sigma_{0}$,
where $\sigma_{0}$ is the surface energy density ($\sigma_{0}\approx16$
eV/nm$^{2}$ for nickel {[}16,17{]}). However, the deformation also
allows a contact area $A_{I}>\Delta A$, that lowers the energy by
$|A_{I}\cdot\sigma_{I}|$, where $\sigma_{I}$ is the metal-carbon
interaction energy density {[}24{]}. 

\begin{figure*}
\begin{centering}
\includegraphics[width=1\textwidth]{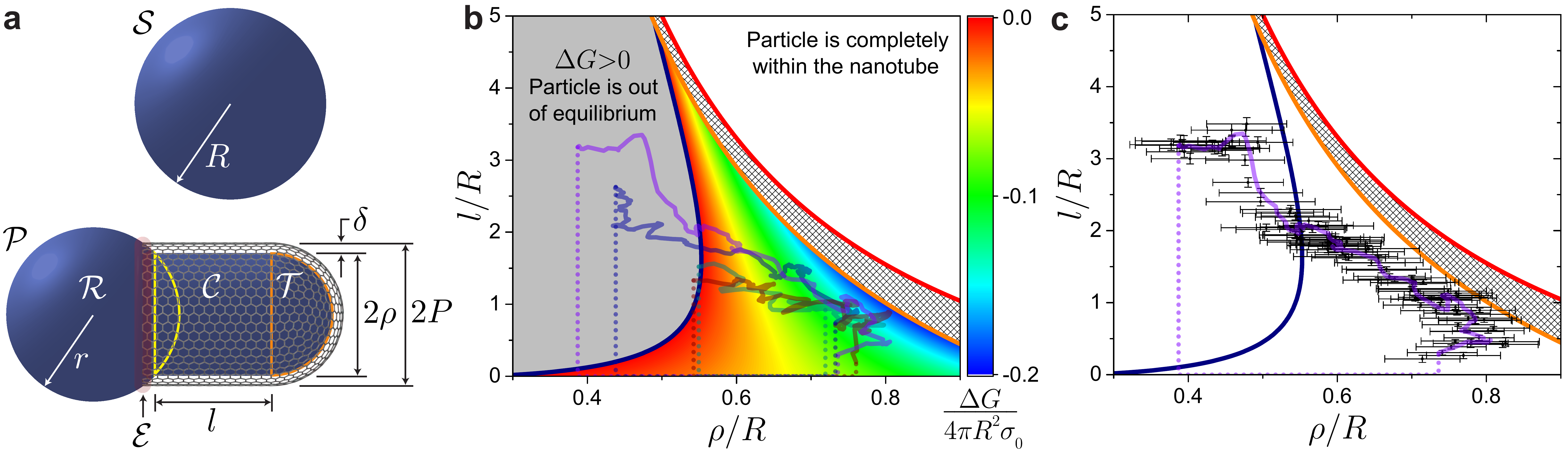}
\par\end{centering}
\caption{\label{fig:Model}Nanoparticle elongation. (a) Before nanotube growth,
the particle is approximately a sphere, $\protect\s$, with radius
$R$. This can elongate into a pear shape, $\protect\p$, with the
growth of the tube. The latter is composed of four regions: a round
region $\protect\r$, an edge region $\protect\e$, a cylindrical
form $\protect\c$, and a tip $\protect\t$ (orange outline) {[}25{]}.
We allow $\protect\c$ to taper with an angle $\theta$, see Fig.
S2, reflecting that the elongated region is often conical, e.g., Fig.
\ref{fig:RepCycle}h, instead of cylindrical, e.g., Fig. \ref{fig:RepCycle}e.
The surface area of this yellow-outlined spherical cap has to be removed
when describing $\protect\r$ as a sphere, as this portion of the
surface is not present. (b) The dimensionless free energy change,
$\Delta G/4\pi R^{2}\sigma_{0}$, versus $\rho/R$ and $l/R$ for
$q=-1/3$ and tapering angle $\theta=0$. The four curves (violet,
royal, cyan, and wine solid lines) are running averages of trajectories
from ETEM measurements, showing the cyclical behavior with different
elongations (and times). When the trajectory goes into the gray, shaded
region, the free energy is positive and the particle is favored to
be outside of the tube (i.e., state $\protect\s$ with a nanotube
attached). In this regime, the particle is certainly out of equilibrium,
as retraction would lower its free energy. However, retraction requires
overcoming an energy barrier and, until $\rho$ and $l$ reach a certain
point, the particle will be trapped in a metastable state. (c) A single
trajectory with both the running average (violet, solid line) and
all the data points (black circles and error bars in both radius and
length) from the ETEM video. Elongation can be quite long, here reaching
three times the radius of the initial particle. The constant radius
and constant length processes are approximately present in these trajectories,
as seen by the step-like features {[}26{]}, see also Fig. \ref{fig:Iso-processes}.
Above the orange solid lines, $\protect\r$ is not sufficiently large
to form a ``capping'' hemisphere {[}27{]}; above the red line, the
tube is large enough to accommodate the entire nanoparticle. The dark
blue line demarcates the equilibrium transition from elongated to
spherical particles for $\theta=0$. Error bars represent plus/minus
one standard deviation. }
\end{figure*}

These give the dominant contributions to the Gibbs free energy change,
$\Delta G$, for the particle going from its spherical form, $\s$,
to the elongated \textquotedblleft pear shape\textquotedblright ,
$\p$,
\[
\Delta G=\Delta A\cdot\sigma_{0}+A_{I}\cdot\sigma_{I}.
\]
This determines whether the particle is thermodynamically favored
to be outside ($\Delta G>0$) or inside ($\Delta G<0$) the carbon
nanostructure. Considering structures consisting of only spherical
and cylindrical regions (Fig. \ref{fig:Model}a), we have $\Delta A=4\pi\left(r^{2}-rh/2-R^{2}\right)+A_{I}$
and $A_{I}=2\pi(l\rho+\rho^{2})$ with $R$ the radius of the initial
spherical nanoparticle, $r$ the radius of the spherical region $\r$
outside the tube, $l$ and $\rho$ the length and radius of the elongated
region $\c$. The nanotube radius is $P=\rho+\delta$, where $\delta\approx0.2$
nm is the gap between the carbon and the metal {[}13,19{]}. The quantity
$h=r-\sqrt{r^{2}-\rho^{2}}$ is the height of the region of overlap
between $\r$ and $\c$ (outlined in yellow in Fig. \ref{fig:Model}a).
When the particle tapers at an angle $\theta$ (i.e., it has a conical
form), these expressions change (see the SI for the full calculation). 

\begin{figure}
\begin{centering}
\includegraphics[width=1\columnwidth]{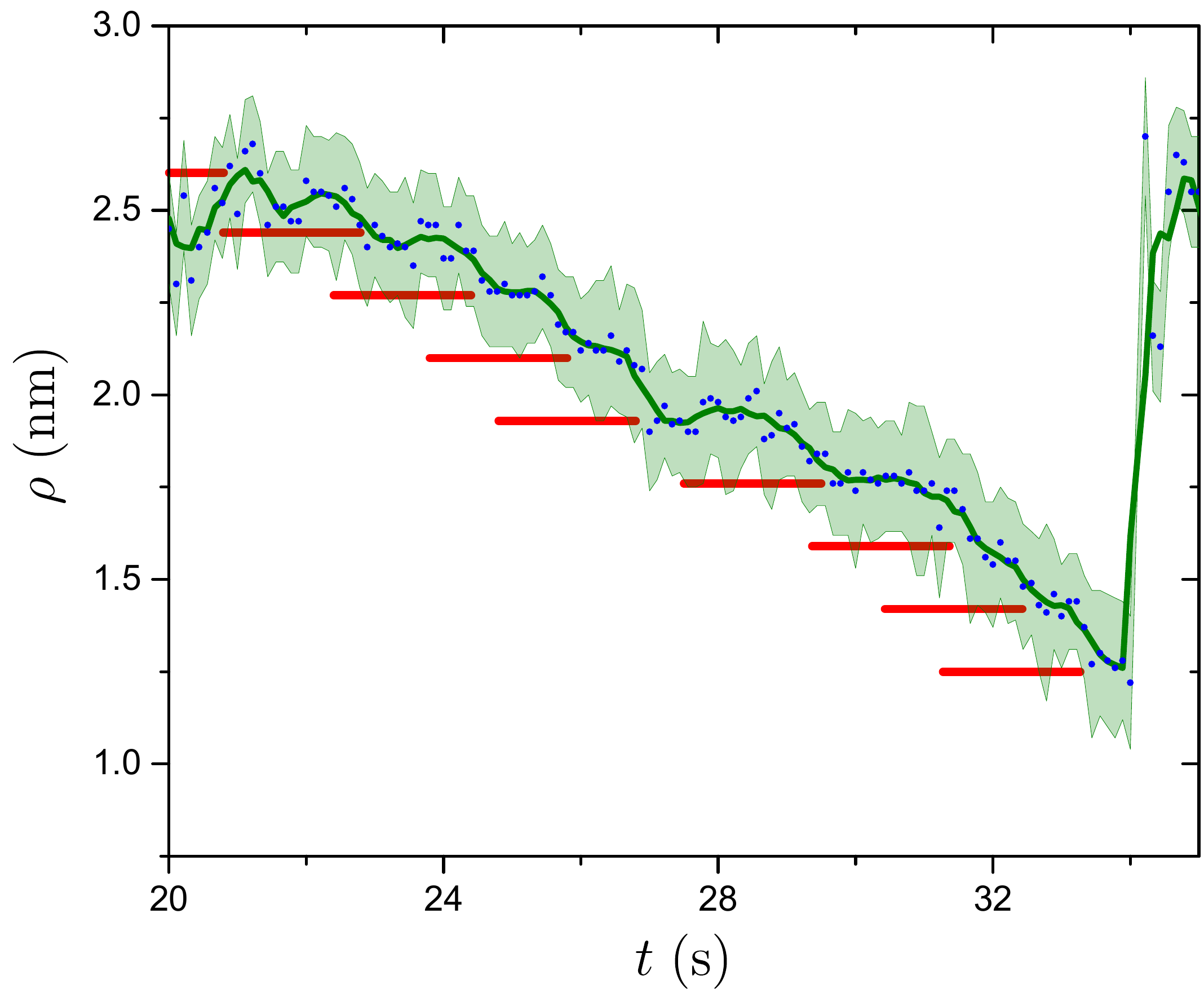}
\par\end{centering}
\caption{Radius of the particle versus time for the cycle in Fig.~\ref{fig:RepCycle}.
The dark green line is the running average of the experimental data
(blue points). As evident by Movie S1, the inner tube nucleation proceeds
first on one side of the particle, then on the other (since the ETEM
provides 2D images, it is unknown what the full 3D process looks like.
For instance, it may proceed by wrapping around the particle). This
process is reflected in a ``stepping down'' of the radius: The red
lines are equally spaced in the vertical direction with spacing $0.34/2=0.17$
nm, i.e., half the radial separation between two consecutive tubes.
The shaded green region is plus/minus one standard deviation. \label{fig:Iso-processes}}
\end{figure}

Figure \ref{fig:RepCycle}i shows the corresponding idealized Gibbs
free energy for Fig. \ref{fig:RepCycle}a-g, which resembles the well-known
Carnot and Otto thermodynamic cycles {[}20{]}. However, here the processes
are approximately iso-radial and iso-longitudinal (i.e., constant
length), as opposed to isothermal, isobaric, etc. (and, as we will
see, also out of equilibrium). Step-like features are visible in the
observed trajectories, Fig. \ref{fig:Model}b,c, especially in radius
versus time, Fig. \ref{fig:Iso-processes}. 

Within the cycle shown in Fig. \ref{fig:RepCycle}i, the particle's
free energy does not always decrease. Depending on the inner tube
radius, the free energy of the particle can be pushed upwards, which
occurs for Fig. \ref{fig:RepCycle}c$\Rightarrow$d onward. In this
case, the inner tube radii are smaller than the threshold value (see
the SI)
\[
P^{\star}\approx(1+q)R+\delta,
\]
where $1+q$ gives an effective (dimensionless) surface energy and
$q=\sigma_{I}/\sigma_{0}$ is the ratio of metal-carbon interaction
to metal surface energy density. For an innermost tube radius below
this threshold, the elongation of the particle requires work, i.e.,
$\partial\Delta G/\partial l>0$ \textendash{} work provided by carbon
addition. 

While carbon addition can serve as an effective driving force for
particle deformation, the free energy eventually becomes positive
with decreasing tube radius and the particle is thermodynamically
favored to be outside of the structure. At this point, the particle
will lower its free energy if it can escape from the tube and restore
its spherical form. This results in a phase diagram for particle morphology
where a critical line \textendash{} determined by the geometry and
interaction parameter $q$ \textendash{} demarcates the regimes where
elongation is favorable and unfavorable, see Fig. \ref{fig:Model}b.
The figure also shows the retracted and elongated structures and the
morphological trajectories ($\rho/R$ and $l/R$ of the particle versus
time) from the ETEM videos. Figures S1 and S2 show the automated data
analysis method and Figs.~S3-S8 give additional schematic and trajectory
information.

Even when it is unfavorable for the particle to be in the tube, it
is still found there actively catalyzing tube growth internally to
the structure. That is, the out-of-equilibrium particle continues
to function, allowing for both longer and more tubes to grow. For
the particle to escape, the particle tip, $\t$, must detach from
the tube cap, which requires overcoming an energy barrier. Thus, nested
tube addition and further elongation occurs until the particle can
transition to a different free energy surface \textendash{} one where
it is not attached to the tube cap \textendash{} and exit the tube.
The energy barrier, though, is substantial. Without surface restructuring
and tapering, the barrier is $2\pi\rho^{2}\cdot\sigma_{I}$ (e.g.,
in Fig. \ref{fig:Model}c greater than 50 eV for most of the cycle)
\textendash{} even though the change in free energy for the complete
retraction is negative. Taking into account optimal restructuring
at $\t$ (i.e., the curvature at the tip decreasing to that at $\r$,
see the SI), the free energy difference (relative to the spherical
form $\s$) during retraction is Eq. (1) with the areas replaced by
$\Delta A^{\pr}=4\pi\left(r^{2}-R^{2}\right)+A_{I}^{\pr}$ and $A_{I}^{\pr}=2\pi l\rho$.
The path f$\Rightarrow$g in Fig. \ref{fig:RepCycle}i shows the free
energy difference as the particle retracts on this alternate free
energy surface, where we include both restructuring and tapering (i.e.,
the conical form of the elongated region, see the SI). However, for
the cycle shown in Fig. \ref{fig:RepCycle} (and Figs. \ref{fig:Model}c
and \ref{fig:Iso-processes}), the tapering angle is only $\theta\approx1{}^{\circ}$
in the highly elongated form. 

We note that the barrier to retraction after the fourth tube is added
is on the order of 1/1000th the total surface free energy ($4\pi R^{2}\sigma_{0}$),
or less than 10 eV. This barrier is easily overcome by further restructuring,
faceting, or other mechanisms, and, in any case, is certainly within
the uncertainties of the data and limitations of the model. The particle
would have to overcome a barrier almost an order of magnitude larger
in order to retract before the fourth tube forms (while more drastic
restructuring, etc., may suppress such a barrier, this process will
be much slower).

\begin{figure*}
\begin{centering}
\includegraphics[width=1\textwidth]{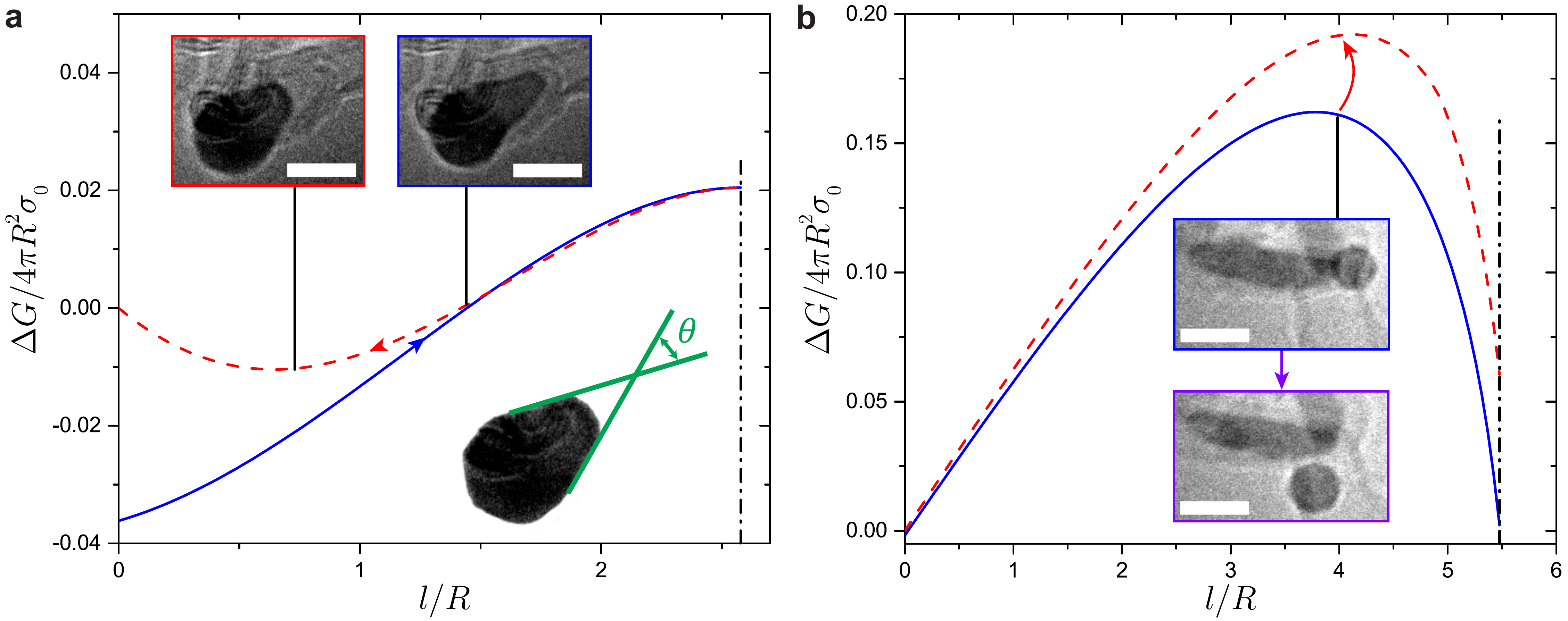}
\par\end{centering}
\caption{\label{fig:Taper}Halted retraction and breakage due to tapered growth.
(a) An inward tapered nanotube grows, elongating the particle with
it (blue, solid line). The growth is performing work on the particle,
but the particle maintains a free energy less than zero and stays
in equilibrium. At a certain length, a free energy surface (dashed,
red line) is available for the particle to partially retract (a full
retraction would increase the free energy). The insets show the particle
and carbon nanostructure at the beginning and end of the retraction
(the points indicated by the black, solid lines) {[}28{]}, as well
as the definition of the tapering angle $\theta$ (bottom right inset).
The elongation and retraction curves converge with each other due
to the inward tapering causing a smaller and smaller contact area
at $\protect\t$ \textendash{} and thus also a lower energy barrier
\textendash{} as $l$ increases (i.e., the radius at the tip of a
tapered particle is $\rho_{0}-l\tan\theta/2$, where $\rho_{0}$ is
the radius where the tube binds at $\protect\e$). Tip restructuring
also lowers the energy barrier, see the SI. (b) A nanotube grows,
eventually resulting in an outward taper, which increases the barrier
to retraction {[}29{]}. The substantial elongation of the particle
is the likely culprit for outward tapering: The tubes bind to steps
on the particle surface {[}21{]}. When the outside portion of the
particle, $\protect\r$, shrinks dramatically, this can decrease the
spatial extent of the steps and contract the radius of the tube, tapering
its end. In this example, the barrier (red, solid arrow) to retract
is about $0.03\cdot4\pi R^{2}\sigma_{0}$ (in the 10\textquoteright s
of eV) even though the tapering angle is only about $-5^{\circ}$.
This prevents the particle from retracting {[}30{]} and the nanotube
drives the breakage of the particle {[}31{]}. The inset images show
the particle before and after breakage {[}32{]}. In both plots, the
black, dash-dotted line shows the maximum extent of $l/R$ for a particle
of that geometry.}
\end{figure*}

Moreover, tapering \textendash{} already discussed above \textendash{}
frequently occurs. Figure \ref{fig:Taper} shows the effect of tapering
on the particle morphology. For inward tapering (Fig. \ref{fig:Taper}a),
the barrier to retraction can be substantially reduced as the particle
tip has a smaller contact area with the carbon nanostructure. Indeed,
the two curves \textendash{} one for elongation and one for retraction
\textendash{} rapidly approach each other (and even cross), indicating
that the particle will detach from the carbon nanostructure at the
tip and retraction will occur. When this occurs in the regime $\Delta G<0$,
the retraction will only be partial, as it is uphill for the particle
to completely exit the nanostructure. There are no thermodynamic forces
that can push the particle up this hill (carbon structure growth can
only perform work outward along the particle's length or, during tube
addition, radially inward). The analytical model we present explains
this behavior quantitatively. Moreover, this demonstrates why the
presence of inward tapering promotes carbon nanofiber formation (over
bamboo or tubular structures): Once a tapered carbon nanostructure
forms, the shorter elongation and incomplete retraction favor the
formation of stacked canonical carbon structures (see Fig. S6). Larger
nanoparticles, for instance, display more tapering (see Fig. S7).
This may be a result of particle curvature \textendash{} the cap forms
to conform to the particle, giving a conical rather than a semi-spherical
cap (kinetics may also play a role here).

Outward tapering (i.e., flaring) has an even more drastic effect.
Figure \ref{fig:Taper}b (also see Movie S2 and Fig.~S8)
shows elongation and retraction curves for an outward tapered particle
($\theta\approx-5^{\circ}$, where the minus sign indicates outward
tapering). The particle can retract when $l$ is small. However, nested
tube addition pushes the particle onto the elongation curve at large
$l$. At this point, there is already a substantial energetic barrier
to transition onto the retraction curve and exit the tube. Thus, even
in the absence of barriers to diffusion, the particle is trapped in
the tube, which eventually encapsulates part of the nanoparticle,
breaking it into two (see the inset in Fig.~\ref{fig:Taper}b).
Such elongated residues have often been observed during ex situ imaging
of the CNT grown in CVD reactors {[}22,23{]} and our model helps understand
this limit on the homogeneity of catalytic products.

In summary, the above model quantitatively captures the morphology
of the catalyst nanoparticles during CNT growth as observed via in
situ ETEM imaging. It is favorable metal-carbon interaction, work
from tube growth, geometry, and the presence of energy barriers that
steer the morphology of the catalytic nanoparticle. While Refs. {[}10{]}
and {[}14{]} discuss the diffusion of nickel atoms, our model demonstrates
the non-equilibrium, cyclical nature of the elongation and retraction,
showing that metastability allows the catalyst to continue growing
the structure. Tapering explains partial retraction and particle length,
helping to understand the formation of fibers and end-product homogeneity.
The model suggests that, for example, altering the value of $q$ (the
ratio of metal-carbon interaction to metal surface energy density)
via, e.g., mixed-metal nanoparticles can encourage certain morphologies
and the growth of particular carbon nanomaterials. Moreover, our results
pose new questions about the role of faceting/surface structure, carbide
formation, the support, rim binding, and kinetics, ones that will
open up novel directions in the investigation and classification of
catalytic behavior. \\

P. A. L. acknowledges support under the Cooperative Research Agreement
between the University of Maryland and the National Institute for
Standards and Technology Center for Nanoscale Science and Technology,
Award 70NANB10H193, through the University of Maryland. We would like
to thank S. Zhu, T. Li, and S. Deffner for helpful discussions.

\onecolumngrid
\vspace{\columnsep}
\vspace{\columnsep}
\vspace{\columnsep}
\twocolumngrid

\begingroup
    \fontsize{9.5pt}{2pt}\selectfont
\begin{enumerate}[label={[\arabic*]},itemsep=-0.8ex,leftmargin=4.7ex]

  \item H. Yan, Q. Li, J. Zhang, and Z. Liu, Carbon \textbf{40},
2693 (2002). 

  \item J. Kong, H. T. Soh, A. M. Cassell, C. F. Quate, and
H. Dai, Nature \textbf{395}, 878 (1998). 

  \item M. F. L. De Volder, S. H. Tawfick, R. H. Baughman,
and A. J. Hart, Science \textbf{339}, 535 (2013). 

  \item A. D. Franklin, Nature \textbf{498}, 443 (2013). 

  \item Z. Han and A. Fina, Prog. Polym. Sci. \textbf{36},
914 (2011). 

  \item M. J. O\textquoteright Connell, Science \textbf{297},
593 (2002). 

  \item S. Ogata and Y. Shibutani, Phys. Rev. B \textbf{68},
165409 (2003). 

  \item S. M. Bachilo, Science \textbf{298}, 2361 (2002). 

  \item B. I. Yakobson and P. Avouris, in Carbon Nanotub.,
edited by M. S. Dresselhaus, G. Dresselhaus, and P. Avouris (Springer
Berlin Heidelberg, Berlin, Heidelberg, 2001), pp. 287\textendash 327. 

  \item S. Helveg, C. López-Cartes, J. Sehested, P. L. Hansen,
B. S. Clausen, J. R. Rostrup-Nielsen, F. Abild-Pedersen, and J. K.
N{\o}rskov, Nature \textbf{427}, 426 (2004). 

  \item H. Amara, C. Bichara, and F. Ducastelle, Phys. Rev.
Lett. \textbf{100}, 056105 (2008). 

  \item Y. Ohta, Y. Okamoto, A. J. Page, S. Irle, and K.
Morokuma, ACS Nano \textbf{3}, 3413 (2009). 

  \item E. Pigos, E. S. Penev, M. A. Ribas, R. Sharma, B.
I. Yakobson, and A. R. Harutyunyan, ACS Nano \textbf{5}, 10096 (2011). 

  \item M. Moseler, F. Cervantes-Sodi, S. Hofmann, G. Csányi,
and A. C. Ferrari, ACS Nano \textbf{4}, 7587 (2010). 

  \item S. B. Vendelbo, C. F. Elkj{\ae}r, H. Falsig, I. Puspitasari,
P. Dona, L. Mele, B. Morana, B. J. Nelissen, R. van Rijn, J. F. Creemer,
P. J. Kooyman, and S. Helveg, Nat. Mater. \textbf{13}, 884 (2014). 

  \item T. A. Roth, Mater. Sci. Eng. \textbf{18}, 183 (1975).

  \item F. R. de Boer, Cohesion in Metals: Transition Metal
Alloys, 2 (North Holland, Amsterdam, 1988).

  \item M. Picher, P. A. Lin, J. L. Gomez-Ballesteros, P.
B. Balbuena, and R. Sharma, Nano Lett. \textbf{14}, 6104 (2014).

  \item P. A. Khomyakov, G. Giovannetti, P. C. Rusu, G. Brocks,
J. van den Brink, and P. J. Kelly, Phys. Rev. B \textbf{79}, 195425
(2009).

  \item H. B. Callen, Thermodynamics and an Introduction
to Thermostatistics, 2nd ed (Wiley, New York, 1985). 

  \item R. Rao, R. Sharma, F. Abild-Pedersen, J. K. N{\o}rskov,
and A. R. Harutyunyan, Sci. Rep. \textbf{4}, 6510 (2014).

  \item Z. He, J.-L. Maurice, A. Gohier, C. S. Lee, D. Pribat,
and C. S. Cojocaru, Chem. Mater. \textbf{23}, 5379 (2011).

  \item J. Gao, J. Zhong, L. Bai, J. Liu, G. Zhao, and X.
Sun, Sci. Rep. \textbf{4}, 3606 (2014). 

  \item Although these particles partially convert to carbide
during growth, the carbon structures (via the inner tube) attach to
the metal terminated surfaces only, as reported for Co-based catalysts
{[}18{]}, and the interaction is not markedly influenced by the internal
particle structure.

  \item We note that only the innermost nanotube is included,
as the outer tubes are expected only to play a secondary role in the
elongated-to-retracted transition. 

  \item We note that other non-idealities can be present but
not visible since the ETEM yields a 2D image.

  \item This line thus delineates where the thermodynamic
expression, Eq. (1), is valid and also where we expect different physical
behavior due to faceting of the particle outside the tube (and potentially
the removal of step edges that anchor the tube, e.g., causing rim
detachment), giving a substantial energetic barrier that will halt
elongation. In Fig. 1c, further elongation would approach this regime,
i.e., where the radius of the elongated region of the particle is
equal to width of the particle outside the tube.

  \item We take $\rho_{0}$ and $\theta$ from immediately
after the retraction using the automated data analysis values. 

  \item This is opposed to inward tapers, which reduce the
barrier.

  \item We note that the particle does not start off on the
elongation curve at $l=0$, but rather nested tube addition pushes
the particle onto this curve at a large value for $l$. The outward
tapered particle might have been able to retract if it was pushed
onto this curve at smaller $l$.

  \item The creation of two new surfaces in the metal particle
of radius $\rho_{0}$ costs $\left(1+q\right)2\pi\rho_{0}^{2}\sigma_{0}$
or about $0.03\cdot4\pi R^{2}\sigma_{0}$ for the small radius neck
shown in the top image. An increased contact of the external, spherical
particle with the carbon nanostructure may actually lower the barrier
to particle breakage. In any case, the tube growth can drive breakage,
but it cannot drive retraction: \emph{There are no external processes
that can assist the exiting of the particle (and the particle can
not follow the blue line for decreasing $l$ as the tube would have
to shrink), but there is a process that assists breakage. }

  \item We take $\rho_{0}$ and $\theta$ from measurements
on the images. 

\end{enumerate}
\endgroup


\clearpage
\newpage

\widetext
\setstretch{1.5}

\renewcommand{\thefootnote}{\fnsymbol{footnote}}

\setcounter{figure}{0}
\renewcommand\thefigure{S\arabic{figure}} 

\setcounter{equation}{0}
\renewcommand\theequation{S\arabic{equation}} 

\renewcommand\thesection{\arabic{section}} 
\renewcommand\thesubsection{\thesection.\arabic{subsection}}

\renewcommand{\bibnumfmt}[1]{[S#1]}
\renewcommand{\citenumfont}[1]{S#1}

\setcounter{table}{0}
\renewcommand\thetable{S\arabic{table}} 

\setcounter{page}{1}
\renewcommand\thepage{S\arabic{page}} 

\begin{center}
{\large\bf Metastable morphological states of catalytic nanoparticles \textendash{}
Supplemental Information} 

\vspace{0.2cm}

Pin Ann Lin,$^{1, 2}$ Bharath Natarajan,$^{3}$ Michael Zwolak,$^{1,\hspace{-3pt}}$~\footnote{michael.zwolak@nist.gov} and Renu Sharma$^{1,\hspace{-3pt}}$~\footnote{renu.sharma@nist.gov} \\

\vspace{0.1cm}

\begingroup
    \fontsize{9.5pt}{8pt}\selectfont
{\em $^{1}$ Center for Nanoscale Science and Technology, National Institute

of Standards and Technology, Gaithersburg, Maryland 20899, USA

$^{2}$ Maryland NanoCenter, University of Maryland, College Park, Maryland, USA.

$^{3}$ Materials Measurement Laboratory, National Institute of Standards and Technology Gaithersburg, MD 20899 USA}

\endgroup

\end{center} 

\setlength{\cftsecindent}{1.2em}
\setlength{\cftsubsecindent}{4.2em}
\setlength{\cftsecnumwidth}{1em}
\setlength{\cftsubsecnumwidth}{2em}

\tableofcontents

\newpage

\section{Environmental transmission electron microscope}

We use an environmental transmission electron microscope (ETEM) [S1]
operated at 200 kV to observe the CNT growth \emph{in situ}. A Ni-SiO$_{x}$
catalyst is dry-loaded onto 200 mesh molybdenum TEM grids. The sample
on the TEM grid is loaded onto a TEM heating holder and introduced
into the ETEM column. The sample is heated to temperatures between
773 K and 798 K in vacuum. After approximately 30 min, C$_{2}$H$_{2}$
is introduced to initiate the CNT growth and a pressure of 0.39 Pa
is maintained during growth. The movies are at a frame rate of 9 $s^{-1}$.
Note that precision of our dimensional measurements is limited by
the pixel resolution ($\approx$0.066 nm) of the images.

\section{Theory of elongation and retraction}

\subsection{Elongation in a cylindrical tube}

When the bulk of the nanoparticle does not change appreciably, i.e., when the
 particle maintains its crystalline structure (as seen
in ETEM observations~[S2]) and ignoring atomic details, the Gibbs
free energy change \textendash{} the catalytic process is at constant temperature
and pressure \textendash{} for elongation with tube growth
will have contributions only from the metal surface energy, the metal-carbon
(surface) interaction, and surface configurations (entropy). The latter
is likely to be small even at the elevated temperatures used for carbon
nanostructure growth. Thus, including only the dominant terms, we
have
\begin{equation}
\Delta G=G_{\p }-G_{\s},\label{eq:G}
\end{equation}
with
\begin{equation}
G_{\s}=\sigma_{0}\cdot4\pi R^{2} \label{eq:Gs}
\end{equation}
and
\begin{equation}
G_{\p}=\sigma_{0}\cdot\left(4\pi r^{2}-2\pi rh\right)+\left(\sigma_{0}+\sigma_{I}\right)\cdot(2\pi l\rho+2\pi\rho^{2}).\label{eq:Gp}
\end{equation}
This gives Eq. (1) in the main text.

In addition, to compute the free energy, we need a volume constraint
on the particle. Since the nickel remains crystalline, its total volume
will be approximately conserved giving the equation
\begin{equation}
\frac{4}{3}\pi R^{3}=\frac{4}{3}\pi r^{3}-\frac{1}{6}\pi h\cdot\left(3\rho^{2}+h^{2}\right)+\pi\rho^{2}l+\frac{2}{3}\pi\rho^{3}. \label{eq:V}
\end{equation}
The left hand side is the volume of the initial, spherical particle.
The terms on the right hand side are, in order, the volume of the
spherical region outside the tube, a correction term that subtracts
the spherical cap of that same region (as the sphere is not complete),
the volume of the cylinder inside the tube, and the volume of the
hemispherical end of the cylinder. Equations~\eqref{eq:G}-~\eqref{eq:V} yield
a set of dimensionless equations in terms of $\rho/R$ and $l/R$
(and $q=\sigma_{I}/\sigma_{0}$) only, which thus characterize the
elongation. This indicates that the energetics of the morphological
changes we examine are scale-invariant, contrary to what is widely
believed~[S3] (these changes will become slower, however, as the
size of the particle increases). Faceting and the density of steps
increase as the curvature increases. Thus, the scale invariance will
be broken, but this is not expected to happen until the particle sizes
are below about 1 nm to 2 nm~[S1].

The first term in Eq.~\eqref{eq:Gp} \textendash{} the one proportional
to only $\sigma_{0}$ \textendash{} is the surface energy of region
$\r$. The second term has both the surface energy and metal-carbon interaction
energy of the regions $\c$ and $\t$. Note that the edge region, $\e$, where
the tube binds to the metal, shown in Fig.~2a of the main text, is
not included in these expressions, as here we are examining the conditions
that determine when the particle will elongate with tube growth and
retract from the tube later. For these processes, the rim region gives
an approximately identical contribution to both the spherical and
``pear'' shapes (to fully understand the origin of tapering, though,
a more detailed treatment of the rim region is necessary). We note
that the outer tubes unbind from the particle during inner tube elongation,
a process that likely has its origins in the kinetics of tube formation
\textendash{} smaller tubes elongate faster as they need less carbon
to grow per unit length. When an outer tube detaches, there will be
an energy penalty. This penalty is less than the gain in free energy
due to the further elongation of the inner tube.

To go further than this descriptive account of the interactions, we
need to express Eq.~\eqref{eq:G}, with the free energies given by Eqs.~\eqref{eq:Gs} and~\eqref{eq:Gp}, in terms of $\rho$ and $l$ only, subjecting
it to the volume constraint, Eq.~\eqref{eq:V}. Using the latter, the
radius of region $\r$ is
\begin{equation}
r=\frac{\rho^{4}+\sqrt[3]{X-Y}+\sqrt[3]{X+Y}}{4C}, \label{eq:r}
\end{equation}
where $C=4R^{3}-2\rho^{3}-3\rho^{2}l$, $X=8\ C^{4}+8\ C^{2}\ \rho^{6}+\rho^{12}$,
and $Y=4\ C\ \sqrt{C^{2}+\rho^{6}}\ \left(2\ C^{2}+\rho^{6}\right)$.
All of the radicands are positive within the region of interest and
thus give real values for $r$ (when $C<0$, i.e., when $l>4\left(R^{3}-\frac{\rho^{3}}{2}\right)/3\rho^{2}$,
gives the region demarcated by the red line in Fig. 3a,b of the main
text). With this value of $r$, we will automatically satisfy the
volume constraint and we can write
\begin{equation}
G(\rho,l).
\end{equation}
That is, the free energy change is a function of only $\rho$ and
$l$ (or, when dimensionless, $\rho/R$ and $l/R$). As seen in Fig.~2 of the main text, the initial particle deformation (large $\rho/R$)
falls within the regime where elongation is thermodynamically favored
($\Delta G<0$). With decreasing $\rho/R$, the morphology transitions into
a regime where elongation is disfavored ($\Delta G>0$).

The free energy for elongation, though, only yields a partial picture.
We also want to know the local thermodynamic forces (the derivative
of the free energy) and the barriers to transition to different free
energy surfaces. Since we have $G(\rho,l)$, the derivative,
\begin{equation}
\frac{\partial\Delta G}{\partial l},
\end{equation}
with respect to length is straight forward to compute, but will yield
unwieldy expressions. Instead, we will work with $G(\rho,r)$ instead
of $G(\rho,l)$. At fixed $\rho$, a change in $r$ gives only a change
in $l$ and vice versa. We can then compute $\partial\Delta G/\partial r\cdot\partial r/\partial l$
and transform back into a function of $\rho$ and $l$. In some sense,
the variables $\rho$ and $r$ are more natural for calculations.
However, $r$ is not the most transparent variable for understanding
elongation. Using Eq.~\eqref{eq:V}, we have
\begin{equation}
l=\frac{4R^{3}-4r^{3}+\ h\cdot(3\rho^{2}+h^{2}\ )/2-2\rho^{3}}{3\rho^{2}},\label{eq:l}
\end{equation}
which allows us to get both $\partial r/\partial l$ and $G(\rho,r)$.
We then find
\begin{equation}
\frac{\partial\Delta G}{\partial l}=\frac{\rho \, r \left(1+q\right)-\rho^{2}}{2r}.\label{eq:dGdl}
\end{equation}
Using Eq.~\eqref{eq:r} for $r$ gives also the desired derivative as
a function of the right arguments. The transition
\begin{equation}
\frac{\partial\Delta G}{\partial l}<0\ \leftrightarrow\ \frac{\partial\Delta G}{\partial l}>0,
\end{equation}
demarcates the region where particle elongation with nanotube growth
will happen spontaneously from that where it requires work. From Eq.~\eqref{eq:dGdl}, the transition line for spontaneous elongation (not to
be confused with the transition line, $\Delta G=0$, which is related
but different) is
\begin{equation}
r^{*}=\frac{\rho}{1+q},\label{eq:rstar}
\end{equation}
where we define $r^{*}$ as the radius of the outer region $\r$ on the
transition line. Thus, the transition line is solely dependent on
the ratio $q=\sigma_{I}/\sigma_{0}$ through an effective (dimensionless)
surface energy density of the particle within the nanotube, $\gamma=1+q$.
With units, this is $(1+q)\sigma_{0}$, which gives the effective
surface energy density of the elongated region in terms of the metal
surface energy attenuated by the metal-carbon interaction. This allows
us to find $l^{*}$ \textendash{} the length of the tube at the transition
line \textendash{} versus $\rho$ by putting Eq.~\eqref{eq:rstar} into
Eq.~\eqref{eq:l}, at
\begin{equation}
\frac{l^{*}}{R}=\frac{1}{3}\left(\frac{4R^{2}}{\rho^{2}}-\frac{2\rho}{\gamma^{3}R}\left(\gamma^{3}+1+\left(1+\gamma^{2}/2\right)\sqrt{1-\gamma^{2}}\right)\right),
\end{equation}
with $\gamma=1+q$ the effective dimensionless surface energy. To
make this expression more transparent, we can solve for the value
of $\rho$ when $l^{*}=0$. This occurs at at
\begin{equation}
\rho^{*}\approx\gamma R=\left(1+q\right)R,
\end{equation}
with corrections that are fourth order in $\gamma$ (in other words,
even for moderate metal-carbon interaction strengths, e.g., $q=-1/3$,
the perturbative expression is accurate). Moreover, in the regime
the experimental measurements are in (i.e., $l$ not very small), the transition line is essentially
\begin{equation}
\frac{l^{*}}{R}\approx\frac{4}{\gamma^{3}}(\gamma R-\rho).
\end{equation}
This means that as $\rho$ gets smaller, the transition that halts spontaneously elongation happens at
longer and longer $l$, drawing out the particle with continued nested tube formation. 

For all results we use $q=-1/3$, which is about half of the value,
$q\approx-6/10$, from DFT calculations of nickel-graphene interaction~[S4]. One expects that the value would be lower for interaction
with nanotubes due to curvature and imperfect contact (i.e., from
additional surface roughness, incommensurate length scales, faceting,
etc.).

\subsection{Retraction from a cylindrical tube}

For retraction to occur, the particle tip has to unbind from the tube.
Direct detachment entails a penalty of $2\pi\rho^{2}\cdot\sigma_{I}$.
This barrier is suppressed by restructuring at the particle tip, $\t$,
and by inward tapering during growth. When the particle tip is not
in contact with the nanotube, it is favorable \textendash{} optimal
in this case \textendash{} for the tip to reduce its curvature to
a sphere of radius $r$ (the radius of the outside region of the particle).
This restructuring reduces the surface energy of the particle. The
free energy of retraction is then
\begin{equation}
G_{\p}^{'}=\sigma_{0}\cdot4\pi r^{2}+\left(\sigma_{0}+\sigma_{I}\right)\cdot2\pi l\rho.
\end{equation}
Volume conservation during retraction is given by
\begin{equation}
\frac{4}{3}\pi R^{3}=\frac{4}{3}\pi r^{3}+\pi\rho^{2}l. \label{eq:Vretract}
\end{equation}
Using Fig.~2a of the main text, the spherical cap (yellow line) is
``cut out'' and moved to the tip (orange). There is exactly a volume
of a sphere of radius $r$ plus the volume of the elongated cylinder.
This restructuring reduces the energy barrier. Since the $r$ before
and after restructuring is different, call them $r_{i}$ and $r_{f}$,
respectively, the energy barrier is
\begin{equation}
\sigma_{0}\cdot4\pi r_{f}^{2}-\left\lbrack (4\pi r_{i}^{2}-2\pi r_{i}h)+\left(\sigma_{0}+\sigma_{I}\right)\cdot2\pi\rho^{2}\right\rbrack ,
\end{equation}
where $r_{i}$ is given by Eq.~\eqref{eq:r} and $r_{f}$ easily found
from Eq.~\eqref{eq:Vretract}.

\subsection{Elongation in an inward tapered tube}

The free energy for elongation with tube growth is
\begin{equation}
G_{\p}=\sigma_{0}\cdot\left(4\pi r^{2}-2\pi rh_{0}\right)+\left(\sigma_{0}+\sigma_{I}\right)\cdot(2\pi\rho_{l}^{2}+\pi\frac{\sqrt{1+s^{2}}}{s}\left(\rho_{0}^{2}-\rho_{l}^{2}\right)),
\end{equation}
where $s=\tan{\theta/2}$ is the slope of the taper with angle $\theta$,
$\rho_{0}$ is the radius at the tube mouth, $h_{0}=r-\sqrt{r^{2}-\rho_{0}^{2}}$,
and $\rho_{l}=\rho_{0}-sl$ is the radius at the tip. The volume constraint
is
\begin{equation}
\frac{4}{3}\pi R^{3}=\frac{4}{3}\pi r^{3}-\frac{1}{6}\pi h_{0}\cdot\left(3\rho_{0}^{2}+h_{0}^{2}\right)+\frac{2}{3}\pi\rho_{l}^{3}+\frac{1}{3s}\pi\left(\rho_{0}^{3}-\rho_{l}^{3}\right).
\end{equation}
The $r$ that satisfies this constraint is given by Eq.~\eqref{eq:r},
but with $C=4-2\ \rho_{l}^{3}-\left(\rho_{0}^{3}-\rho_{l}^{3}\right)/s$.

\subsection{Retraction from an inward tapered tube}

As with the cylindrical tube, when the tip detaches from the tube,
the surface will restructure. For an inward tapered tube, the radius
of the tip that minimizes the free energy is $r$ (the radius of the
outside portion of the particle). The free energy for retraction is
\begin{equation}
G_{\p}^{'}=\sigma_{0}\cdot\left(4\pi r^{2}-2\pi rh_{0}+2\pi rh_{l}\right)+\left(\sigma_{0}+\sigma_{I}\right)\cdot\pi\frac{\sqrt{1+s^{2}}}{s}\left(\rho_{0}^{2}-\rho_{l}^{2}\right),
\end{equation}
where $h_{l}=r-\sqrt{r^{2}-\rho_{l}^{2}}$ (unlike the cylindrical
case, the spherical caps do not have the same height and they do not
cancel). The volume constraint is
\begin{equation}
\frac{4}{3}\pi R^{3}=\frac{4}{3}\pi r^{3}-\frac{1}{6}\pi h_{0}\cdot\left(3\rho_{0}^{2}+h_{0}^{2}\right)+\frac{1}{6}\pi h_{l}\cdot\left(3\rho_{l}^{2}+h_{l}^{2}\right)+\frac{1}{3s}\pi\left(\rho_{0}^{3}-\rho_{l}^{3}\right).
\end{equation}
These equations are used to plot the retraction from the inward tapered
tube in Fig.~4a of the main text.

\subsection{Elongation in an outward tapered tube}

The free energy for elongation with tube growth is
\begin{equation}
G_{\p}=\sigma_{0}\cdot\left(4\pi r^{2}-2\pi rh_{0}\right)+\left(\sigma_{0}+\sigma_{I}\right)\cdot(2\pi\rho_{l}^{2}+\pi\frac{\sqrt{1+s^{2}}}{s}\left(\rho_{l}^{2}-\rho_{0}^{2}\right)),
\end{equation}
where $s=\tan{\left|\theta\right|/2}$ is the slope of the taper with
angle $\theta$, $\rho_{0}$ is the radius at the tube mouth, $h_{0}=r-\sqrt{r^{2}-\rho_{0}^{2}}$,
and $\rho_{l}=\rho_{0}+sl$ is the radius at the tip. The volume constraint
is
\begin{equation}
\frac{4}{3}\pi R^{3}=\frac{4}{3}\pi r^{3}-\frac{1}{6}\pi h_{0}\cdot\left(3\rho_{0}^{2}+h_{0}^{2}\right)+\frac{2}{3}\pi\rho_{l}^{3}+\frac{1}{3s}\pi\left(\rho_{l}^{3}-\rho_{0}^{3}\right).
\end{equation}
The $r$ that satisfies this constraint is given by Eq.~\eqref{eq:r},
but with $C=4-2\ \rho_{l}^{3}-\left(\rho_{l}^{3}-\rho_{0}^{3}\right)/s$.

\subsection{Retraction from an outward tapered tube}

As with the cylindrical tube, when the tip detaches from the tube,
the surface will restructure. For an outward tapered tube, the radius
of the tip that minimizes the free energy is $\max{(r,\rho_{l})}$.
When $\rho_{l}<r$, the free energy for retraction is
\begin{equation}
G_{\p}^{'}=\sigma_{0}\cdot\left(4\pi r^{2}-2\pi rh_{0}+2\pi rh_{l}\right)+\left(\sigma_{0}+\sigma_{I}\right)\cdot\pi\frac{\sqrt{1+s^{2}}}{s}\left(\rho_{l}^{2}-\rho_{0}^{2}\right),
\end{equation}
where $h_{l}=r-\sqrt{r^{2}-\rho_{l}^{2}}$. The volume constraint
is
\begin{equation}
\frac{4}{3}\pi R^{3}=\frac{4}{3}\pi r^{3}-\frac{1}{6}\pi h_{0}\cdot\left(3\rho_{0}^{2}+h_{0}^{2}\right)+\frac{1}{6}\pi h_{l}\cdot\left(3\rho_{l}^{2}+h_{l}^{2}\right)+\frac{1}{3s}\pi\left(\rho_{l}^{3}-\rho_{0}^{3}\right).
\end{equation}
When $\rho_{l}>r$, the free energy for retraction is
\begin{equation}
G_{\p}^{'}=\sigma_{0}\cdot\left(4\pi r^{2}-2\pi rh_{0}+2\pi\rho_{l}^{2}\right)+\left(\sigma_{0}+\sigma_{I}\right)\cdot\pi\frac{\sqrt{1+s^{2}}}{s}\left(\rho_{l}^{2}-\rho_{0}^{2}\right).
\end{equation}
The volume constraint is
\begin{equation}
\frac{4}{3}\pi R^{3}=\frac{4}{3}\pi r^{3}-\frac{1}{6}\pi h_{0}\cdot\left(3\rho_{0}^{2}+h_{0}^{2}\right)+\frac{2}{3}\pi\rho_{l}^{3}+\frac{1}{3s}\pi\left(\rho_{l}^{3}-\rho_{0}^{3}\right).
\end{equation}
These equations are used to plot the retraction from the outward tapered
tube in Fig.~4b of the main text.

\clearpage
\section{Image Processing}

\begin{figure*}[h]
\begin{centering}
\includegraphics[width=1\textwidth]{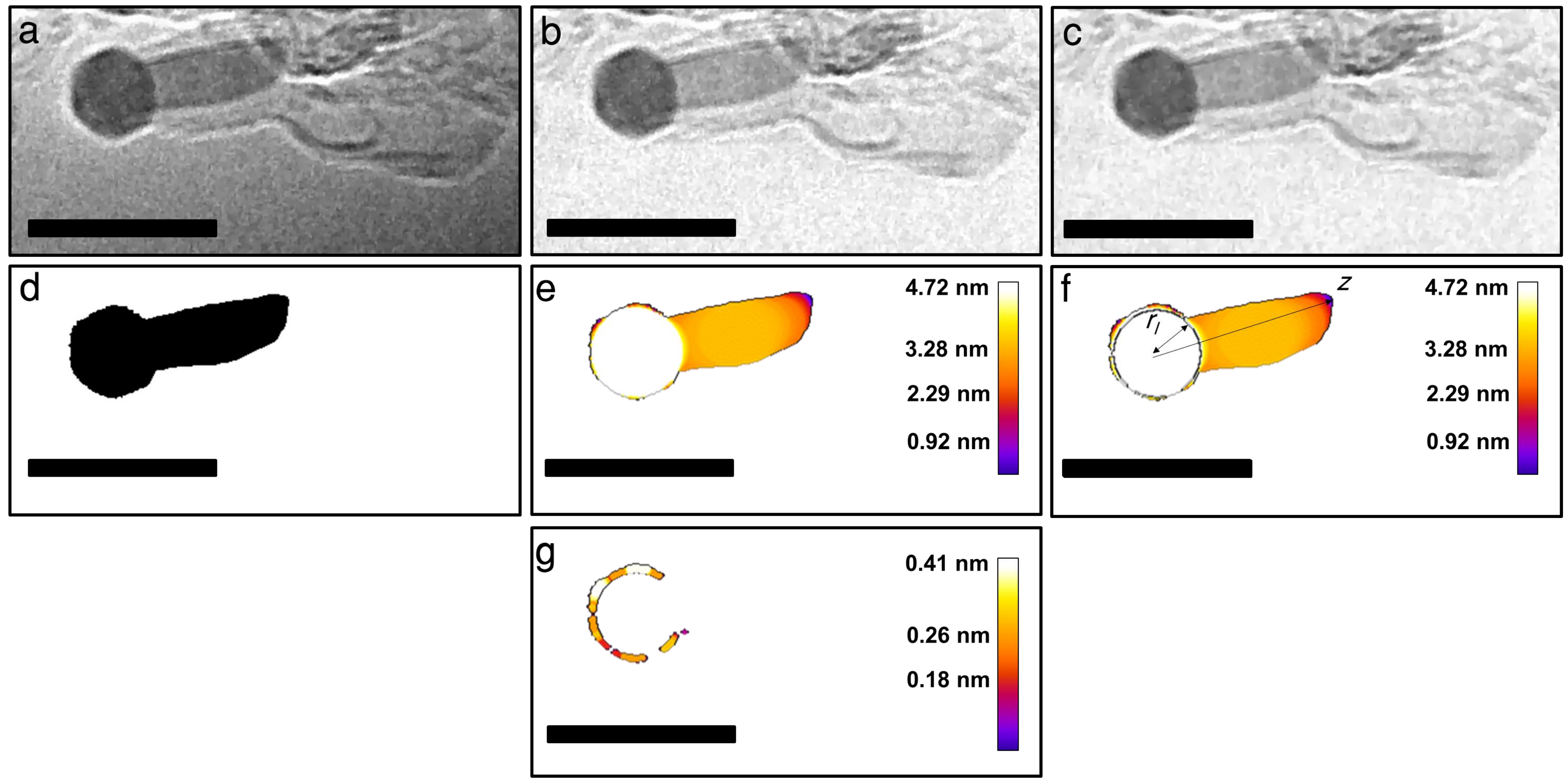}
\par\end{centering}
\caption{\label{fig:ImageAnalysis1}Image processing from a real-time video of BCNT
growth. (a) Representative time slice from Movie~\ref{fig:ImageAnalysis1}; (b) background
subtraction applied to (a); (c) anisotropic diffusion smoothing
applied to (b); (d) binarization applied to (c); (e)
Visual representation of the local thickness calculation applied to
(d), with the thickness labeled according to the color map
inset in the image; (f) Image (e) with in-circle radius $r_{L}$
and loci of in-circle centers $z$ indicated; (g) The local thickness
map of the fringe regions of the spherical domain of the particle.
Scale bars are 10 nm.}
\end{figure*}

A sum total of 579 frames require analysis for the extraction of morphology
descriptors defined in the main text. To address this need, we develop
an algorithm that performs accurate, unbiased binarization of the
image series. Figure~\ref{fig:ImageAnalysis1} shows the image processing steps applied to
a representative image (Fig.~\ref{fig:ImageAnalysis1}a) of an elongated particle. A background subtraction
removes low frequency non-uniformities in intensity (Fig.~\ref{fig:ImageAnalysis1}b). An
``anisotropic diffusion'' (Perona-Malik diffusion) smoothing then
reduces image noise (Fig.~\ref{fig:ImageAnalysis1}c). This smoothing technique preserves
edges, lines, and finer details important for image interpretation.
This image is then thresholded to an appropriate intensity. The method
then isolates the remaining noisy objects by size (units of pixels squared)
and eliminates them using image subtraction to get the final
binarized image.

\begin{figure*}
\begin{centering}
\includegraphics[width=1\textwidth]{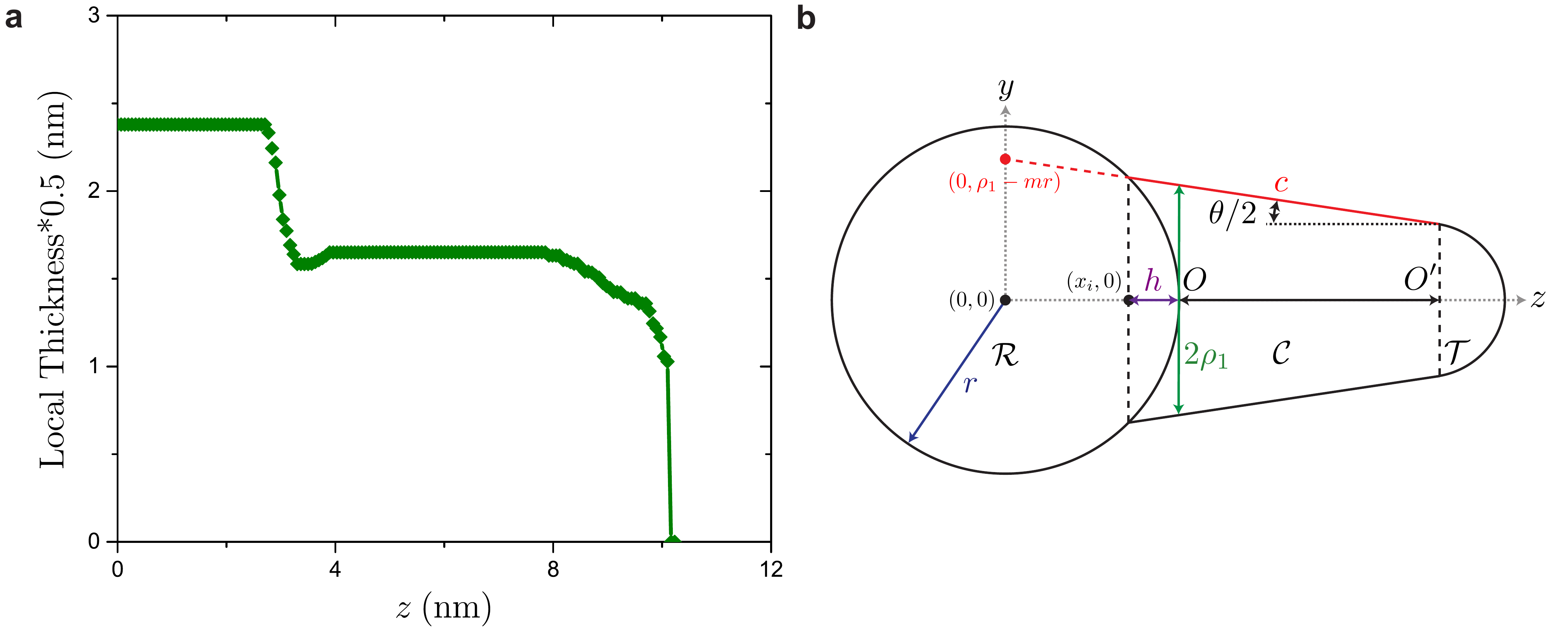}
\par\end{centering}
\caption{\label{fig:ImageAnalysis2}Example data and model. (a) The local thickness
profile along the loci of the center of the fit in-circles ($z$-axis). This is for the same structure as in Fig.~\ref{fig:ImageAnalysis1}. 
(b) Schematic showing the model structure and the parameters for the
computation of $h$ (note that we designate this height $h_0$ when the structure is tapered). The point of intersection of the line $c$ [with slope $m$ and intercept $(0,\rho_{1}-mr)$] with the circle $\r$
(centered at the origin with a radius $r$) is calculated. The $x$
coordinate of this point subtracted from $r$ gives $h$. The values
$m$, $r$, and $\rho_{1}$ are measured using the analysis 
in Fig.~\ref{fig:ImageAnalysis1}.}
\end{figure*}

We compute the local thickness of the particle at each pixel of the
binarized image (Fig.~\ref{fig:ImageAnalysis1}e), defined as follows~[S6,S7]. For $\Omega$ the set of all points in the particle and
${\overset{}{p}}_{1}$ an arbitrary point in the particle, the local
thickness, $\rho({\overset{}{p}}_{1})$, is the largest circle that
contains the point and is completely within the particle's boundary,
\begin{equation}
\rho\left({\overset{}{p}}_{1}\right)=2\ \max\left(\left\{ \sigma|{\overset{}{p}}_{1}\in \mathrm{cir}({\overset{}{p}}_{2},\sigma)\subseteq\Omega,{\overset{}{p}}_{2}\in\Omega\right\} \right).
\end{equation}
Here, $\mathrm{cir}({\overset{}{p}}_{2},\sigma)$ is the set of points inside
a circle with center ${\overset{}{p}}_{2}$ and radius $\sigma$.
From the local thickness map, we also find the radius of the largest
circle, $r_{L}$, in the image (Fig.~\ref{fig:ImageAnalysis1}f), subtract this circle from
the image, and obtain the average of the local thickness of the remaining
fringe region. The radius is of the outer region, $r$, is the sum
of $r_{L}$ and the fringe thickness. The standard deviation in the
fringe thickness is taken to be the uncertainty in $r$ (Fig.~\ref{fig:ImageAnalysis1}g).
We then extract the local thickness profile along the loci of the
center of these fit circles ($z$-axis in Fig.~\ref{fig:ImageAnalysis1}f and Fig.~\ref{fig:ImageAnalysis2}a) from
each image, as well as the taper angle $\theta$ (from the slope of
the thickness profile).

The length \emph{l} is the sum of $OO^\pr$ and $h$ (in Fig.~\ref{fig:ImageAnalysis2}b), where
$OO^\pr$ is the distance between points $O$ and $O^\pr$. The point $O^\pr$
is the center of the circle fit to the tip of the elongated portion
and $h$ is the point of intersection of line $c$ and the circle
$R$ (in Fig.~\ref{fig:ImageAnalysis2}b). The slope of the line $c$ is $\tan\theta$. The
intercept is the thickness at point $E$ ($\rho_{1}$), the taper
$\theta$, and the radius of the outer circle $r$. For an outer circle
centered at the origin, the points of intersection come from the equations
\begin{equation}
x^{2}+y^{2}=r^{2}
\end{equation}
and 
\begin{equation}
y=x\tan\theta+\frac{\rho_{1}}{2}-r\tan\theta.
\end{equation}
We subtract the $x$ coordinate of the point of intersection ($x_{i}$)
from $r$ to obtain $h$. The average radius of the elongated region
($\rho$) is the mean thickness value at each point along the $z$-axis
between points ($x_{i}$,0) and $O'$ in the image. The uncertainty
in $\rho$ is the standard deviation in the thickness of the fringe
elements of the elongated region, i.e., the edge variations outside
of the fit circles.

\newpage
\clearpage
\section{Additional Data and Discussion}

\begin{figure*}[h]
\begin{centering}
\includegraphics[width=1\textwidth]{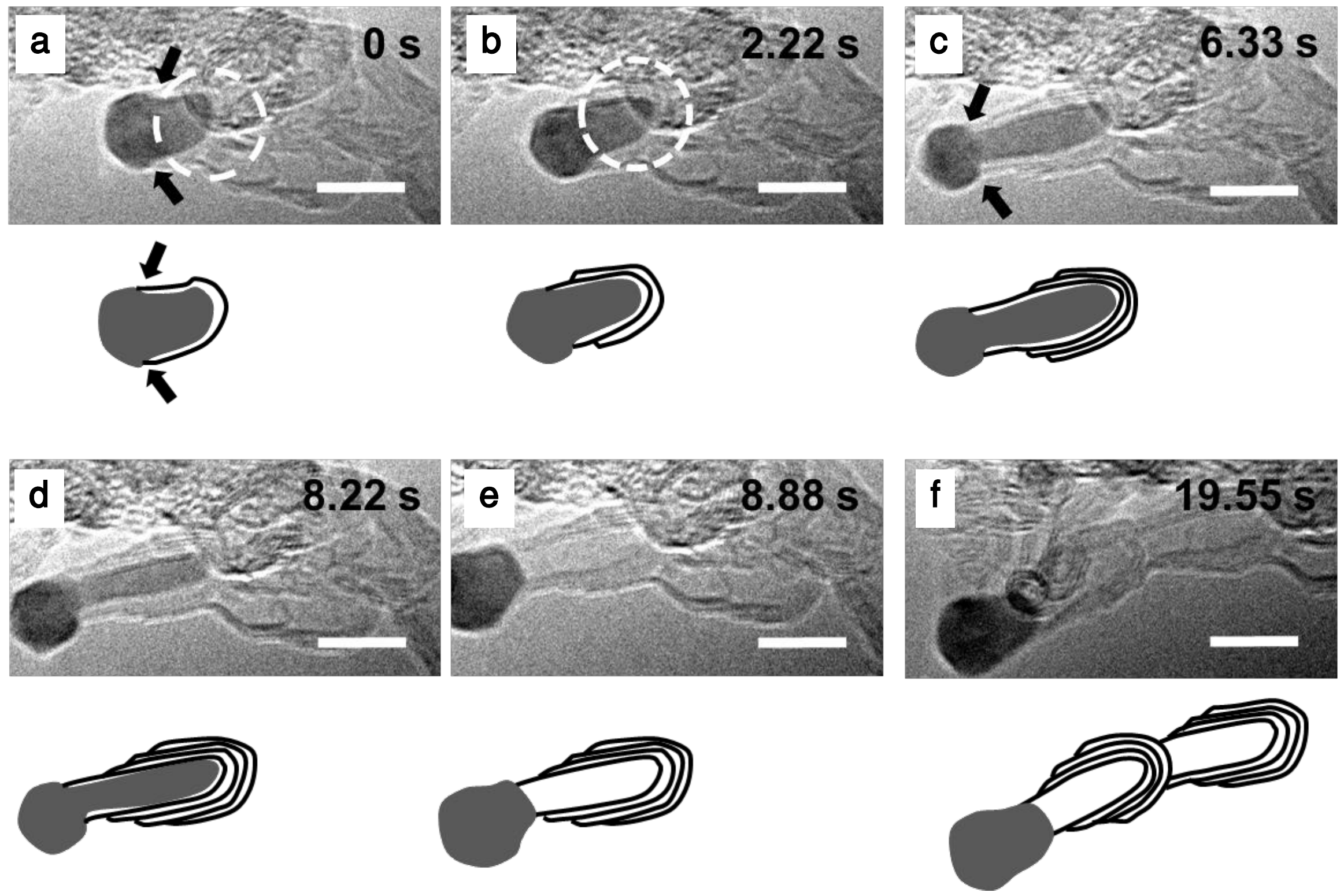}
\par\end{centering}
\caption{\label{fig:CycleSchematic}Snapshots from a real-time video of bamboo-like
carbon nanotubes (BCNT) growth. A portion of the nickel catalyst nanoparticle
($R=(3.2\ \pm\ 0.1)$ nm in radius) elongates inside the tubular structure
during growth. When the radius of the inner tube reaches $P=(1.4\ \pm\ 0.2)$
nm [i.e., $P=(1.2\ \pm\ 0.2)$] \textendash{} less than half the radius of the initial nickel particle
\textendash{} the nanoparticle exits the tube and recovers its spherical
shape. Schematic drawings below each frame illustrate the process,
where the solid shape depicts the catalyst nanoparticle and lines
the CNT. Scale bars are 5 nm. The video sequence shows: (a,b) The
tubes anchor to step edges (pointed to by black arrows) and the tube
caps attach to the upper half of the catalyst particle (indicated
by white circles) during both the early stages of growth and elongation.
(c) New steps form at the interface between the upper and lower halves
of the particle, which results in a more defined interface (pointed
to by black arrows). These steps provide energetically favorable sites
for new nanotubes to nucleate~[S5], always with a conical cap,
from inside the original tube with a consistent ($0.34 \pm 0.08$)
nm spacing. (d) Outer tubes detach from the particle but stay in contact
with newly-formed inner tubes. (e) After the particle detaches from
the cap of the inner most tube, it recovers a roughly spherical shape.
(f) In the original shape, a new hemispherical carbon cap forms with
the rim anchoring at the surface steps on the particle. The elongation
process then repeats.}
\end{figure*}

Figure~\ref{fig:CycleSchematic} shows a series of snapshots from a real-time video of bamboo
carbon nanotube (BCNT) growth recorded after the growth has started
from a $R=(3.2\ \pm\ 0.1)$ nm radius Ni catalyst particle. Schematic
drawings below each frame illustrate the observed changes in catalyst
morphology and the BCNTs formation process. The sequence here starts
from when the catalyst nanoparticle has changed to a pear-like shape
with the CNTs anchored at well-defined step edges (marked by arrows
in Fig.~\ref{fig:CycleSchematic}a). At this point, the particle is starting its elongation.
During this process, new tubes are added into the interior of the
carbon structure. At the end of the approximately 6 s of elongation
(Fig.~\ref{fig:CycleSchematic}a-c), the elongated particle has a length that is approximately
three times the radius of the original particle. The outer CNT then
detaches from the lower half of the particle and the particle roughly
recovers its original, spherical form with the innermost tube's rim
still attached to the steps (Fig.~\ref{fig:CycleSchematic}e). BCNTs form through the cycles
of nanoparticle elongation and retraction, which occur with frequencies
in the range 0.013 s\textsuperscript{-1} to 0.086 s\textsuperscript{-1}

\begin{figure*}[h]
\begin{centering}
\includegraphics[width=0.6\textwidth]{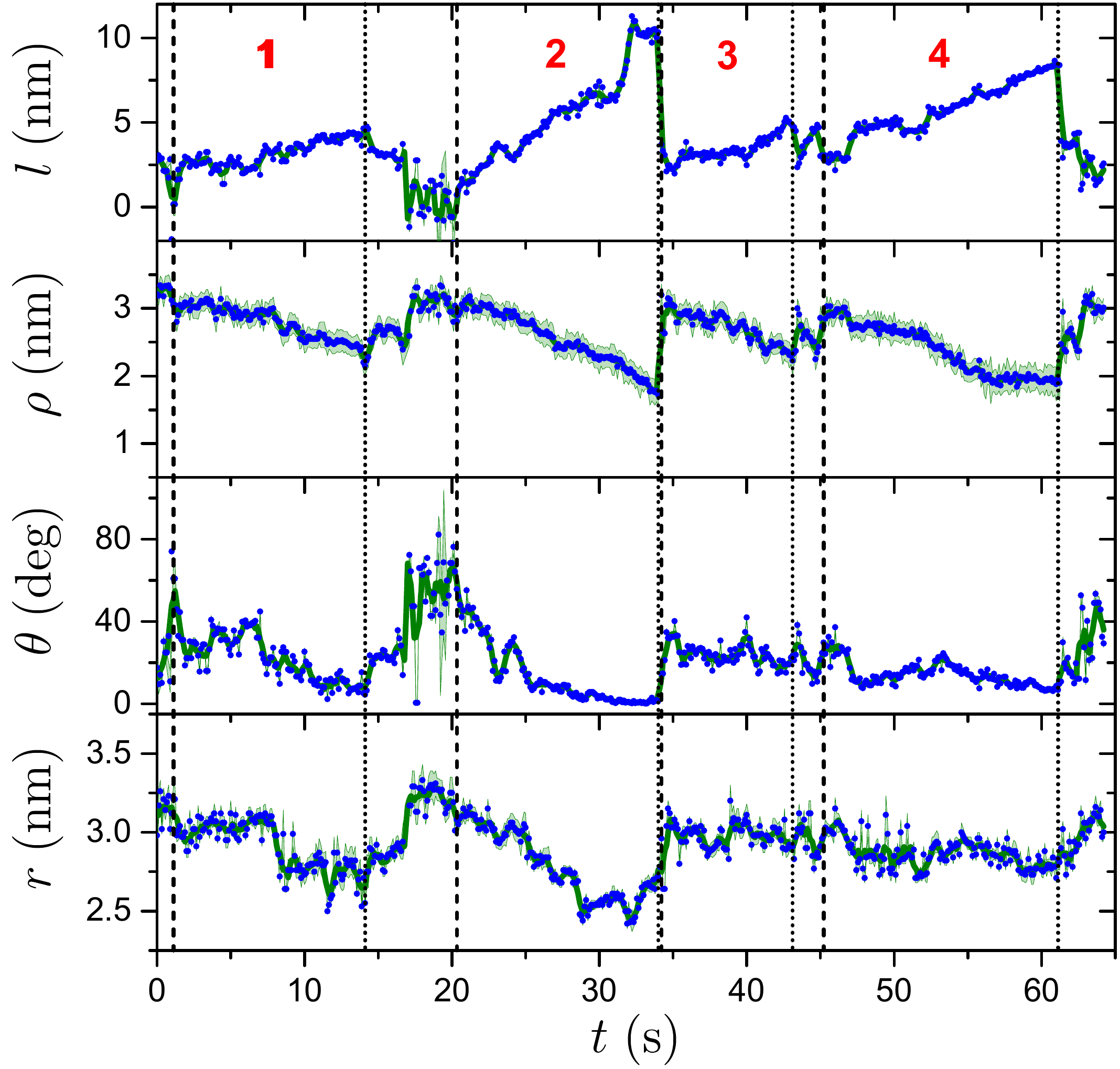}
\par\end{centering}
\caption{\label{fig:StructuralData} Structural data from the automated extraction.
These panels show the time traces of the analyzed video. The four
cycles are taken from this data and are marked by the vertical lines
(dashed lines indicate the beginning and dotted lines the end of the
cycle). The red numbers label the cycle. Figure 2c of the main text
shows cycle 2. Figure S5a,b,c show cycles 1, 3, and 4, respectively.
In addition to the main cycles, there are some shorter elongation
and partial retraction events. Dark green lines indicate the running
average, the shaded green region represents plus/minus one standard
deviation, and the blue circles are the data points.}
\end{figure*}

\begin{figure*}
\begin{centering}
\includegraphics[width=1\textwidth]{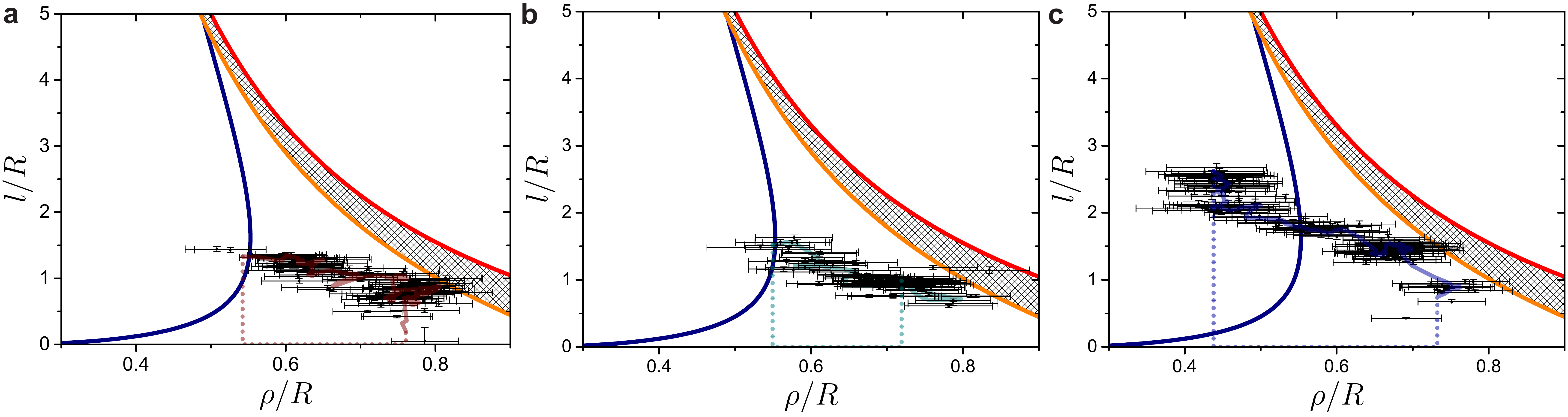}
\par\end{centering}
\caption{\label{fig:AddCycles}Three additional cycles of elongation and retraction.
(a-c) These are the same plot as Fig. 2c of the main text except they are for
the three other cycles shown in Fig. 2b of the main text. The carbon
nanostructures are more highly tapered for these cycles, resulting
in retraction at larger values of $\rho/R$. The blue line indicates
$\Delta G=0$ for the case $\theta=0$ (i.e., no tapering). Above
the red and orange lines are the regions where the particle is completely
in the tube and where the particle would require additional faceting/restricting
to elongate, respectively. The latter, in particular, would act as
a barrier to further elongation, which is seen from these trajectories
\textendash{} they ``avoid'' the orange boundary line. Error bars
are plus/minus one standard deviation.}
\end{figure*}

\begin{figure*}[h]
\begin{centering}
\includegraphics[width=0.6\textwidth]{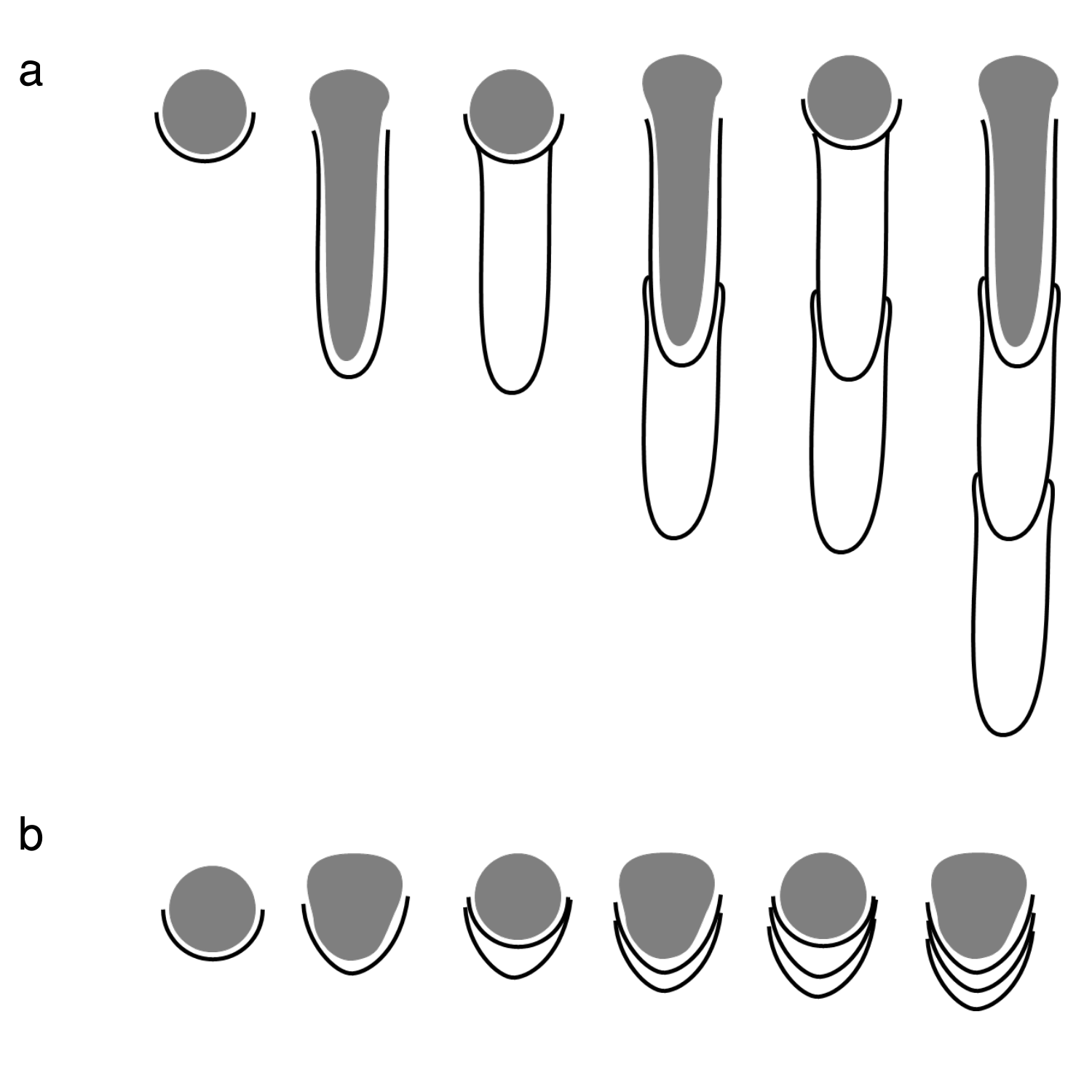}
\par\end{centering}
\caption{\label{fig:Schematic}Schematic of carbon nanostructure formation. (a)
BCNTs form when the degree of particle tapering is small. (b) CNFs
form when the degree of particle tapering is large, as the latter
encourages retraction to be partial and to occur at much smaller elongation
lengths, see Fig.~4a of the main text.}
\end{figure*}

\begin{figure*}[h]
\begin{centering}
\includegraphics[width=1\textwidth]{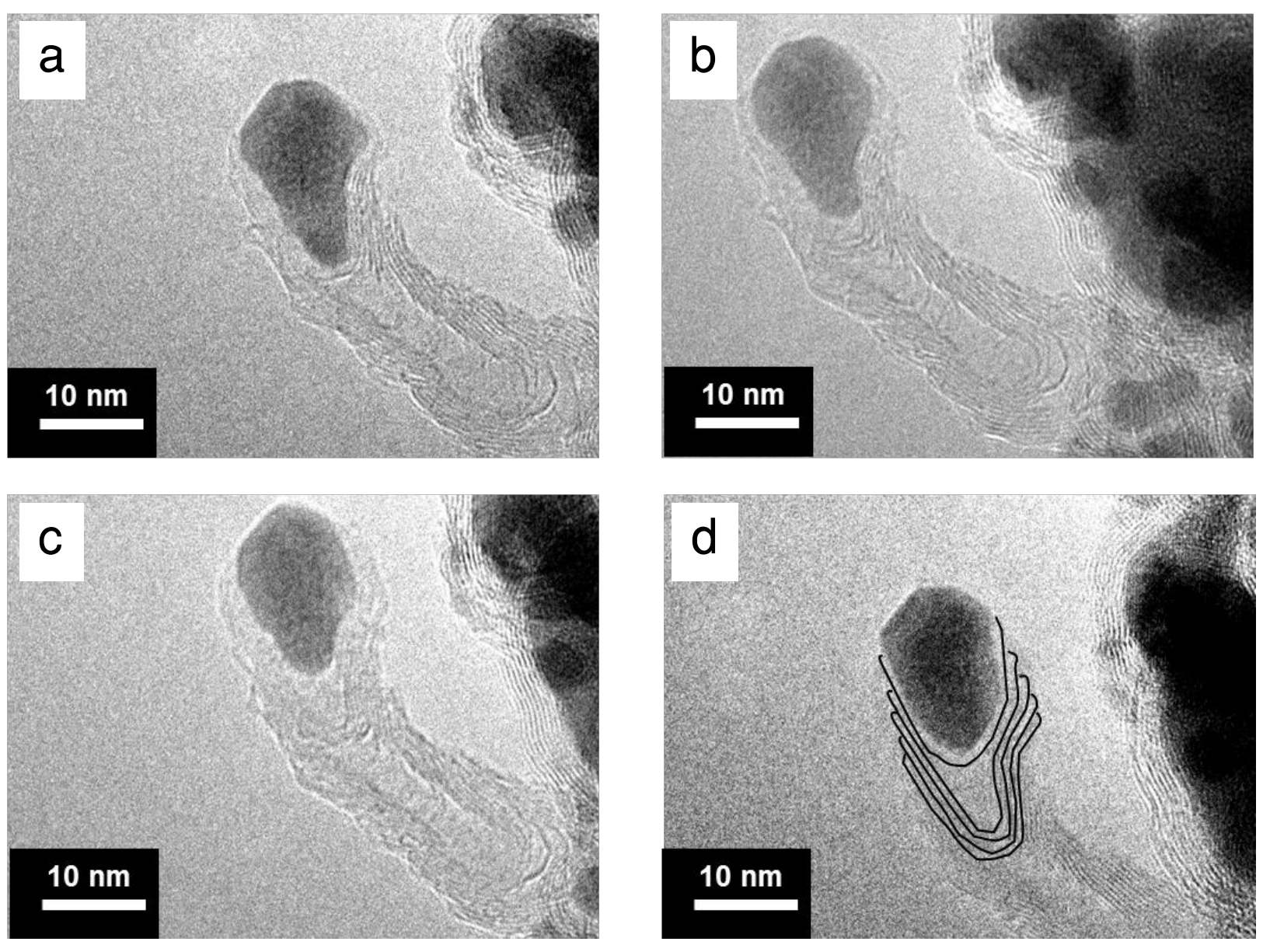}
\par\end{centering}
\caption{\label{fig:Frames}Frames from a real-time video of CNF growth. The
$\approx9$ nm diameter Ni nanoparticle catalyst shape nearly unchanged
during the CNF growth. Scale bars are 10 nm.}
\end{figure*}

Figures~\ref{fig:StructuralData} and~\ref{fig:AddCycles} show data extracted from the ETEM videos.
 Figure~\ref{fig:Schematic} shows a schematic of how carbon nanostructure growth proceeds
for non-tapered and tapered catalytic nanoparticles. Figure~\ref{fig:Frames} is a series of frames from a real-time video of a carbon
nanofiber (CNF) growth from an approximately \emph{R} = 4.5 nm Ni
catalyst nanoparticle at the tip. The video sequence starts when the
catalyst particle is partially elongated inside a CNF (Fig.~\ref{fig:Frames}a).
The degree of the particle elongation is ($0.25 \pm 0.04$) (the
average ratio of elongation to particle diameter), which is much shorter
elongation than the smaller particle analyzed in the main text. The
higher degree of tapering, possibly due to how the carbon cap forms,
results in the shorter elongation and in CNF growth rather than BCNT
growth. A more detailed analysis of the binding region and the interplay
with surface energies is required to confirm that cap formation is
indeed the mechanism that drives the higher degree of tapering. Kinetics
also can play a role here, as larger tubes (that form on the larger
nanoparticle) have a smaller variation in the rate at which they grow.
When a small nested tube forms, its growth rate significantly surpasses
the previous tubes growth rate, causing detachment of that larger
tube. This is less likely to occur when the tube radius is large.
Importantly, the scale invariance of the model in the main text demonstrates
that the smaller shape changes are not due to unfavorable energetics
(except potentially for the energetics/structure of rim binding).

\begin{figure*}[h]
\begin{centering}
\includegraphics[width=1\textwidth]{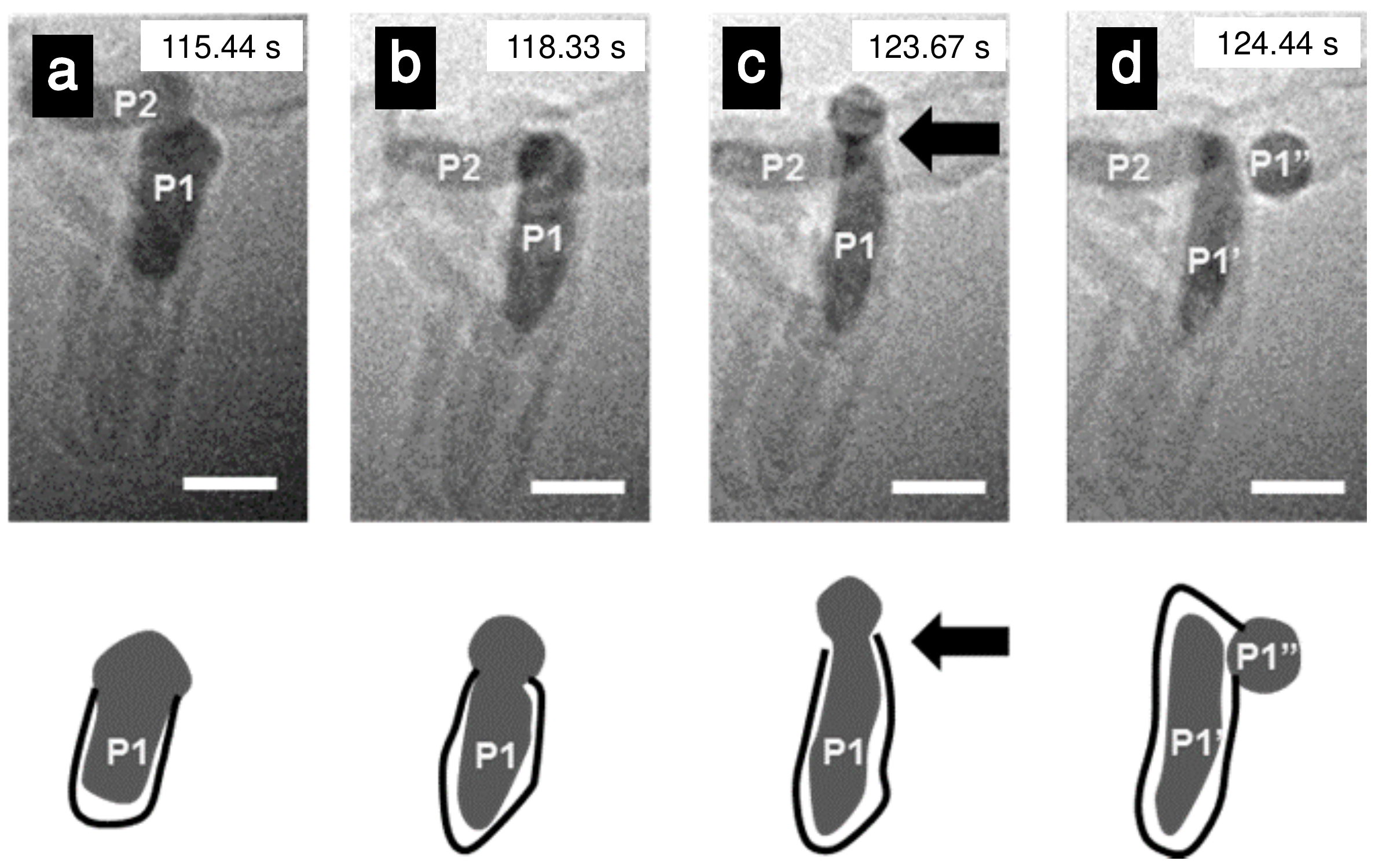}
\par\end{centering}
\caption{\label{fig:Snapshots}Snapshots from a real-time video of CNT growth.
The catalyst nanoparticle (P1) elongates and then breaks into two
parts (P1' and P1'') during growth. Below each frame, the schematic
shows the catalyst (solid shape) and CNT (lines). Scale bars are 5
nm.}
\end{figure*}

Figure~\ref{fig:Snapshots} shows a series of images extracted from the video showing
an occurrence of positive tapering during elongation. The free energy
landscape for this breakage event is in Fig. 4b of the main text.

\newpage

\newpage

\newpage

\newpage

\newpage

\end{document}